\documentclass[%
 reprint,
 amsmath,
 amssymb,
 aps,
 prd,
 superscriptaddress,
 floatfix,
 longbibliography,
 noeprint
 ]{revtex4-2}
 
\usepackage{orcidlink}
\usepackage{graphicx}
\usepackage{float}
\usepackage{hyperref}\hypersetup{hidelinks,colorlinks=false}
\usepackage{placeins}
\usepackage{silence}

\def\ET{E_{\scriptstyle\textrm{T}}}

\begin{document}

\begin{flushright}{\large PITT-PACC-2505-v1}\end{flushright}

\title{Ring-based ML calibration with \textit{in situ} pileup correction for real-time jet triggers}

\author{Benjamin T. Carlson\,\orcidlink{0000-0002-7550-7821}}\thanks{\href{mailto:bcarlson@westmont.edu}{bcarlson@westmont.edu}}
\affiliation{Department of Physics and Engineering, Westmont College, Santa Barbara CA, 93108}
\affiliation{Department of Physics and Astronomy, University of Pittsburgh, Pittsburgh PA, 15260}

\author{Stephen T. Roche\,\orcidlink{0000-0002-3878-5873}}\thanks{\href{mailto:stephen.roche@health.slu.edu}{stephen.roche@health.slu.edu}}
\affiliation{School of Medicine, Saint Louis University, St.\ Louis MO, 63103}
\affiliation{Department of Physics and Astronomy, University of Pittsburgh, Pittsburgh PA, 15260}

\author{Michael Hemmett\,\orcidlink{0009-0007-9810-8800}}
\affiliation{Department of Physics and Engineering, Westmont College, Santa Barbara CA, 93108}

\author{Tae Min Hong\,\orcidlink{0000-0001-7834-328X}}\thanks{\href{mailto:tmhong@pitt.edu}{tmhong@pitt.edu}}
\affiliation{Department of Physics and Astronomy, University of Pittsburgh, Pittsburgh PA, 15260}

\date{\today}

\begin{abstract}
\noindent
We present a machine learning (ML) method to calibrate hadronic jet energy in real-time trigger systems of the High-Luminosity Large Hadron Collider (HL-LHC) using an efficient implementation on field programmable gate arrays (FPGA). Regression is done to estimate the transverse energy of jet candidates, using concentric rings of electromagnetic and hadronic contributions in $0.1{\times}0.1$ towers around fixed-radius cone jet seeds, that accounts for \textit{in situ} pileup correction. Classification separates hard-scatter jets from those due to pileup using the same inputs; its output provides a correction for the regression estimate. The algorithm is tested on simulated samples using an ATLAS-inspired detector in the dense environment of 200 simultaneous proton-proton collisions per bunch crossing. Our method improves the signal efficiency of saving Higgs pair production in $H\!H{\rightarrow}b\bar{b}b\bar{b}$ by a factor of two over the traditional cone jet algorithm in real-time trigger systems.
\end{abstract}

\maketitle

\textit{Introduction.}---Precise determination of hadronic jet energy is essential for the flagship physics goals at CERN's High-Luminosity Large Hadron Collider (HL-LHC) \cite{HLLHC}, one of which is the study of the Higgs potential with hadronic final states \cite{DiVita:2017eyz,Baglio:2012np,Dawson:2015oha,Baur:2002rb,Plehn:2005nk,Djouadi:1999ei}. Improved calibration in the real-time trigger system that decides whether to keep or discard the events would lower the trigger threshold. This, in turn, increases the yield of events that contain Higgs boson pairs that decay to $b$-quarks, \textit{i.e.}, $H\!H{\rightarrow}b\bar{b}b\bar{b}$ ($H\!H_{4b}$), which improves the experimental sensitivity to the Higgs potential through self-coupling \cite{ATLAS:2022jtk,HDBS,ATLAS:2018rvj,CMS:2018sxu}.

Calibration of jet energy from a quark or gluon that emerges from the hadronization process is challenging in the dense environment of the $\langle\mu\rangle{\,=\,}200$ simultaneous proton-proton collisions per bunch crossing at the HL-LHC. The current state-of-the-art algorithm used in the software trigger and offline reconstruction is to run a sequence beginning with particle flow that connects the tracker and calorimeter systems~\cite{ATLAS:2024xna,ATLAS:2017ghe,CMS:2017yfk}, followed by constituent  subtraction that are used as input to jet finding using the anti-$k_t$ algorithm \cite{Cacciari:2008gp,Berta:2014eza,Berta:2014eza,Berta:2019hnj,Cacciari:2014gra,Komiske:2017ubm,CMS:2020ebo,ATLAS;2017csk}. The jet objects are corrected by average pileup in proportion to the jet area~\cite{Cacciari:2007fd,Cacciari:2008gn}. Finally, calibration and pileup rejection taggers are performed on the jet candidates \cite{ATLAS:2023tyv,ATLAS:2020cli,CMS:2011jes,CMS:2016lmd,ATLAS:2015ull,ATLAS:2014jvt}. 

The above sequence is the standard algorithm for offline data analyses, but the experiments at the HL-LHC do not have sufficient computing resources at the real-time trigger systems, which need efficient implementations that can be evaluated in fixed latency of $\mathcal{O}(1)\,\mu$s \cite{Khachatryan:2016bia,Sirunyan:2020zal,Achenbach:2008zzb,CERN-LHCC-2017-020,CERN-LHCC-2020-004}. In particular, the underlying anti-$k_t$ algorithm may add challenges in implementing on field programmable gate arrays (FPGA) due to its iterative design. Typically, simpler fixed-radius cone-based algorithms have been used in real-time systems~\cite{Bystricky:2003pc,Mehdiyev:1999xfa}.  Alternative approaches~\cite{Schlag:2670301,Odagiu:2024bkp} build on advances in machine learning (ML) for FPGA, including neural networks~\cite{Duarte:2018ite,Aarrestad:2021zos} and boosted decision trees (BDT)~\cite{Hong:2021snb,Carlson:2022dgb,Serhiayenka:2024han,Summers:2020xiy}. Our approach is to take cone jets as inputs to perform FPGA-optimized classification and regression.

\begin{figure}[b!]
\centering%
\includegraphics[width=0.48\columnwidth]{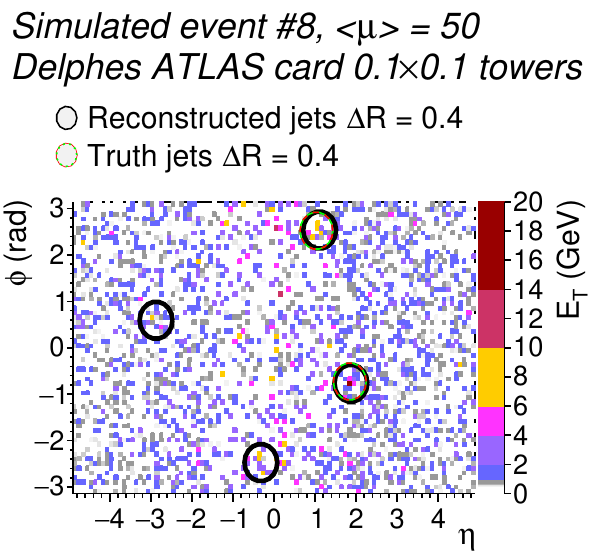}%
\includegraphics[width=0.48\columnwidth]{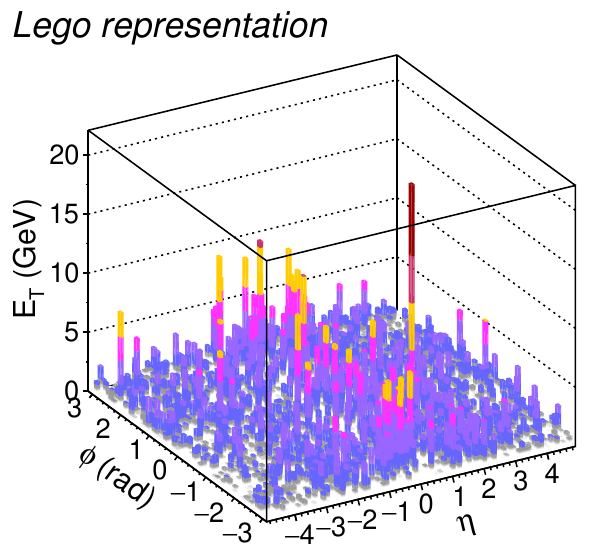}\\
\includegraphics[width=0.48\columnwidth]{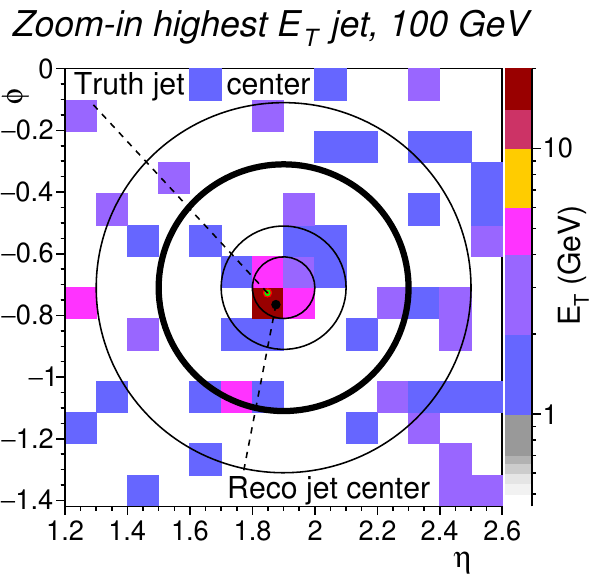}%
\includegraphics[width=0.48\columnwidth]{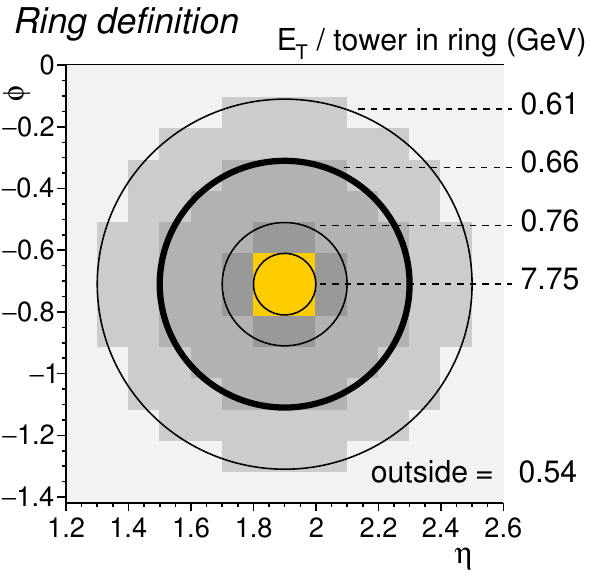}%
\caption{
  Simulated dijet event. Top row: Tower $\ET$ with 4 truth jets (circles) and 2 of them matching anti-$k_t$ jets with $\eta{\,>\,}0$. $\ET^\textrm{jet}{\,>\,}20\,$GeV is applied. Bottom row: $100\,$GeV jet, its $\ET^\textrm{tower}$ (left), and ring$_\textrm{1-4}$ definition and $\langle\ET^\textrm{tower}\rangle_\textrm{ring}$ (right). The region \textit{outside} is for the $0.7{\times}0.7$ square outside ring$_4$. 
  \label{fig:display}
}
\end{figure} 

\textit{Setup.}---We created simulated samples using publicly available Monte Carlo software. Two samples were made \cite{Roche:2025}, each with 100k $pp$ collision events: dijet events for the primary background and $H\!H_{4b}$ for the signal. Simultaneous $pp$ collisions were simulated by overlaying each event with a number of pileup events generated separately with $\langle\mu\rangle{\,=\,}200$ \cite{pileup-atlas}, corresponding to the expected levels at Run 4 and 5 of HL-LHC (2030-2041) \cite{Apollinari:2015wtw,run45}. (Samples of $\langle\mu\rangle{\,=\,}50$ were also made to cross check.) The physics processes were generated with MadGraph5 3.5.0~\cite{Alwall:2011uj}. The generation of pileup events, showering, and hadronization were done by Pythia8~\cite{Sjostrand:2014zea}. Detector simulation and event reconstruction were performed by Delphes 3.4 using the ATLAS card with pileup~\cite{deFavereau:2013fsa}. One example dijet event at $\langle\mu\rangle{\,=\,}50$ is shown on the top row of Fig.~\ref{fig:display}; two generator-level ``truth jets'' and four anti-$k_t$ reconstructed jets are identified; the bottom row zooms-in on the $100\,$GeV jet with rings defined later.

Electromagnetic (\textsc{em}) and hadronic (\textsc{had}) calorimeter energies are used as inputs to estimate the jet energy projected to the plane transverse to the beam direction ($\ET$). The $\eta$-$\phi$ resolution of the inputs is $0.1{\times}0.1$, like the ATLAS Trigger Towers in Run 2 (2015--2018) \cite{Aad:2008zzm} and jFEX Towers in Run 3 (2022--2026) \cite{Aad:1602235}. The sample of reconstructed jets are identified with the anti-$k_t$ clustering algorithm with a minimum $\ET{\,>\,}20\,$GeV and a radius parameter $R{\,=\,}0.4$ with FastJet~\cite{Cacciari:2011ma} using the particle flow (PFlow) energy reconstruction algorithm with pileup (PU) suppression.  In this paper, these are referred to as ``offline jets'' as they are similar to jets in the software-based algorithms for offline processing as well as for the High Level Trigger systems \cite{ATLAS:2024xna}.

\begin{figure}[tbp!]
\centering\includegraphics[width=0.995\columnwidth]{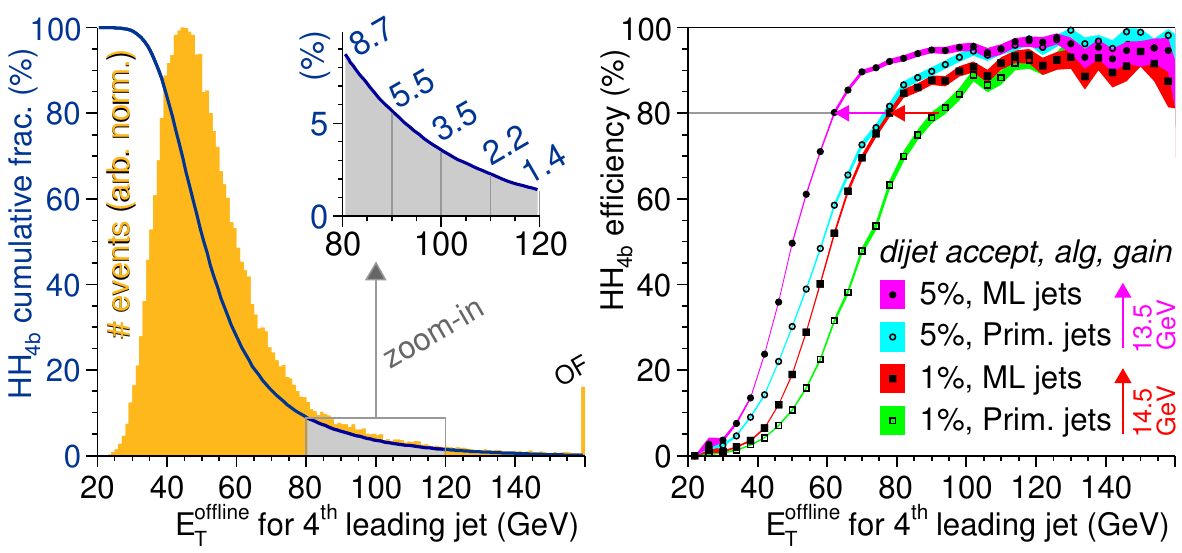}
\caption{
  Performance of ML jets. Left: Distribution of $4^\textrm{th}$ leading jet $\ET^\textrm{offline}$ (histogram); cumulative fraction of $H\!H_{4b}$ above the $x$-axis threshold (curve). The inset shows the $80$-$120\,$GeV region with values. Right: Equal-rate $H\!H_{4b}$ efficiency vs.\ $4^\textrm{th}$ leading jet $\ET^\textrm{offline}$ for $5\%$  ($\circ$) and $1\%$ ({\tiny$\square$}) acceptance of dijet events.
  \label{fig:turnon}
}
\end{figure}

Our approach is to consider a jet candidate after they are created at the real-time trigger system and apply our ML on it to calibrate the jet $\ET$. To mock-up a realistic scenario, we deploy a sliding window algorithm on calorimeter towers, inspired by ATLAS~\cite{Mehdiyev:1999xfa}, to create a set of  ``primitive jets'' \cite{sliding_window}. If a primitive jet is within $\Delta R{\,<\,}0.3$ of any offline jet, where $(\Delta R)^2{\,=\,}(\Delta\eta)^2{+}(\Delta\phi)^2$, it is considered ``matched'' to a ``HS jet'' from the hard-scatter (HS) process. At the target anti-$k_t$ jet's $\ET$ value of $60$-$100\,$GeV---the range where the turn-on feature of the efficiency curve activates (discussed later in Fig.~\ref{fig:turnon})---$70$-$90\%$ of primitive jets are matched (discussed later in Fig~\ref{fig:roc}). The reference $\ET$ value of the HS jet is the anti-$k_t$ $\ET$ value of the corresponding offline jet, which is motivated by the assumption that the best trigger performance will be achieved by matching to the offline energy scale as closely as possible. Otherwise, if no match is found, the primitive jet is considered a pileup jet, or ``PU jet.''

\textit{Machine learning.}---The ML is composed of two BDT models, BDT$_\textrm{ET}$ and BDT$_\textrm{HS}$, whose outputs are combined at the end. The former performs regression as the ``$\ET$ estimator'' and the latter performs classification as a ``Hard Scatter tagger.'' Both models use the same set of input variables from calorimeters. There are sixteen quantities: $\eta$ of jet of interest and sets of three $\ET$ values ($\ET^\textsc{em}$, $\ET^\textsc{had}$, $\ET^\textsc{sum}$) for each of the four rings around the jet shown in Fig.~\ref{fig:display} and defined next, where \textsc{sum} is of the electromagnetic and hadronic components, as well as the set for the smallest three rings combined as ring$_\textrm{jet}$. Each ring is defined by the sum of $0.1{\times}0.1$ tower $\ET$ within the annuli around the $\eta$-$\phi$ center of the jet defined by the following boundaries. The number of $0.1{\times}0.1$ towers are given in parentheses with the ring labels: $\Delta R{\,<\,}0.1$ (4 towers in ring$_\textrm{1}$), $0.1{\,\le\,}\Delta R{\,<\,}0.2$ (12 towers in ring$_\textrm{2}$), $0.2{\,\le\,}\Delta R{\,<\,}0.4$ (40 towers in ring$_\textrm{3}$), and $0.4{\,\le\,}\Delta R{\,<\,}0.6$ (68 towers in ring$_\textrm{4}$). When we refer to primitive jets, they correspond to $\Delta R{\,<\,}0.4$ (56 towers in ring$_\textrm{jet}$). Due to their square nature, a tower is included in a given ring if its $\eta$-$\phi$ center is within the annulus definition. Visualization of a $100\,$GeV jet with rings are shown in Fig.~\ref{fig:display}.

\begin{figure}[b]
\centering\includegraphics[width=0.995\columnwidth]{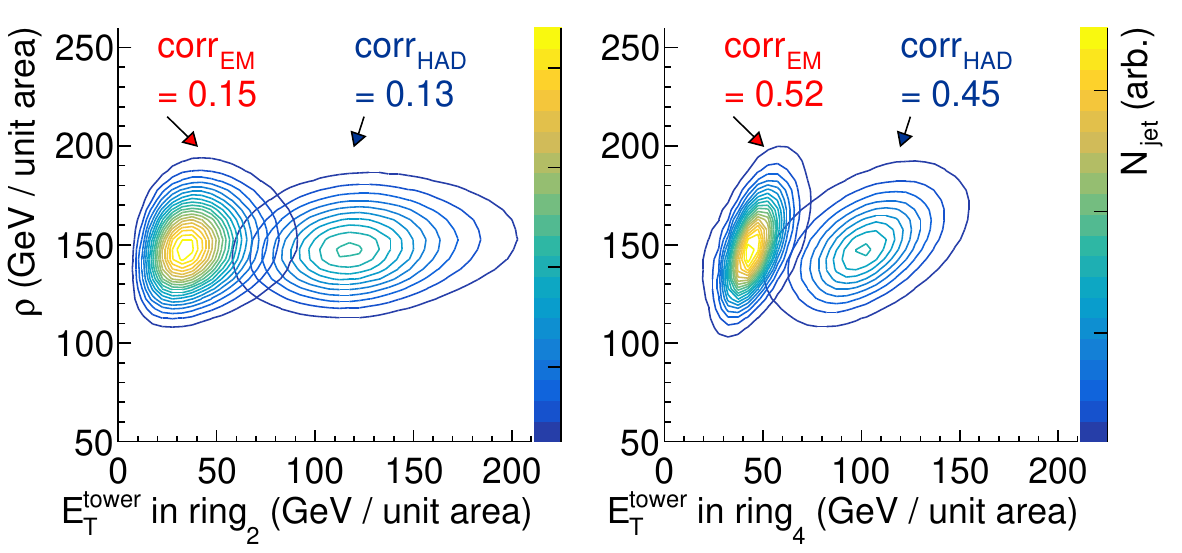}
\caption{
  Correlations between the event $\ET$ density $\rho$ and tower $\ET$ normalized to unit area in $\eta$-$\phi$ for ring$_2$ (left) and ring$_4$ (right) defined in the text. In each plot, both the contours of EM and hadronic components are shown separately as well as their correlation coefficients.
  \label{fig:rho}
}
\end{figure}

The ring definitions are guided by the physics process. The most probable $\ET$ distribution for a HS jet is highly energetic with concentration of $\ET$ near the jet center with lower values fading away. In contrast, a PU jet is the sum of stochastic energy deposits so their distributions are not as strongly peaked. Following this logic, the rings are defined so that ring$_\textrm{jet}$ is the $\ET$ of the primitive jet; ring$_\textrm{1-3}$ are its three concentric decomposition. The two outermost ring$_\textrm{3,4}$ is designed to measure the local pileup level around the jet. On the one hand, we do not expect ring$_\textrm{1,2}$ to be correlated to the ambient $\ET$ density of the due to pileup processes. On the other hand, we expect ring$_\textrm{3,4}$ to be strongly correlated to the ambient $\ET$ density and serve as the input variable to provide \textit{in situ} pileup suppression. To compare the pileup density in each ring, we use the pre-defined pileup density computed as the median of calorimeter tower energies, which we denote as $\rho$~\cite{Cacciari:2007fd}. The median suppresses the upward bias of the mean due to the contribution of jets from the hard scatter process.  Figure~\ref{fig:rho} confirms our expectations by showing the lack of correlation of ring$_2$ and the strong correlation of ring$_4$ to $\rho$. The correlation coefficients for ring$_2$ and ring$_4$ are $0.15$ ($0.13$) and $0.52$ ($0.45$), respectively, for $\textsc{em}$ ($\textsc{had}$).

\begin{figure}[t]
\centering\includegraphics[width=0.995\columnwidth]{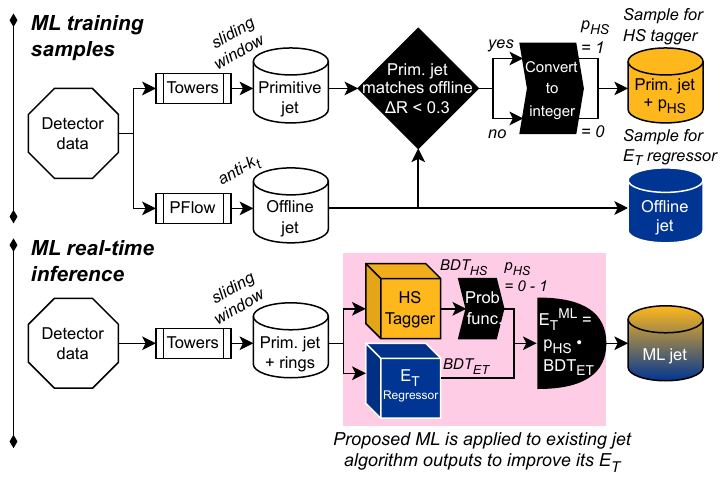}
\caption{
  Flowchart for obtaining the simulated training sample (top) and the real-time inference of incoming data (bottom). The pink box encloses the ML proposed in this paper.
  \label{fig:flowchart}
}
\end{figure}

ML training is done using {\textit{T}\textsc{MVA}} \cite{Hocker:2007ht}. Both BDT$_\textrm{ET}$ and BDT$_\textrm{HS}$ are each composed of 30 trees with a maximum depth of 8 using adaptive boosting. Half the sample is randomly assigned to training, with the rest assigned to testing. The workflow to obtain the training sample is shown on the top half of Fig.~\ref{fig:flowchart} and described below. 

BDT$_\textrm{ET}$ is regression trained to estimate the $\ET$ given a primitive jet. It is trained on the sample of offline jets, using only the calorimeter-derived features above as inputs, with the offline $\ET$ calculated by anti-$k_t$ as the target value. As the $\ET$ spectrum of these jets is exponentially decreasing, weighting is applied to effectively flatten the $\ET$ distribution to avoid biasing the BDT in favor of lower $\ET$. Events are weighted by the inverse of the population of the corresponding bin in the $\ET$ distribution. It is important to note that because the BDT$_\textrm{ET}$ is trained on anti-$k_t$ offline jets, it is optimized assuming that any jet under consideration is from hard scatter. We will see that this is a reasonable assumption and that a correction is made for the assumption later in the paper.

\begin{figure}[t]
\centering\includegraphics[width=0.995\columnwidth]{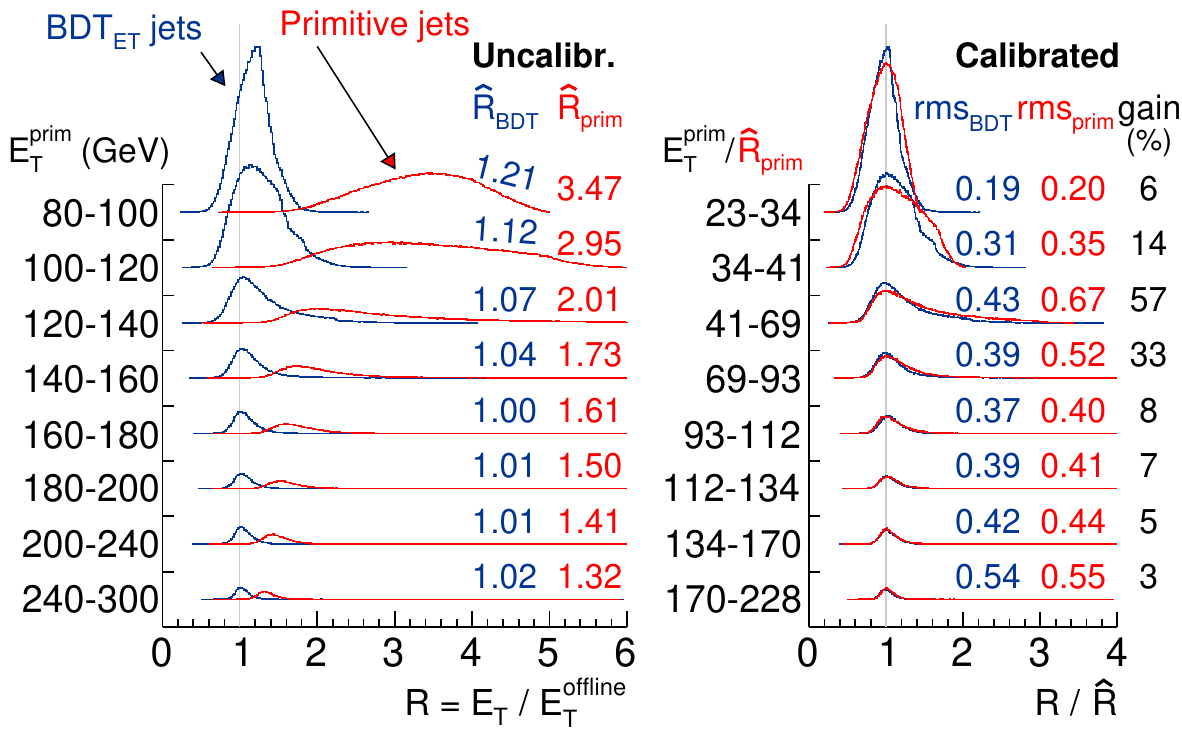}%
\caption{
  Performance of BDT$_\textrm{ET}$. Left: Distribution of uncalibrated $\ET$ resolutions in eight slices of $\ET^\textrm{prim}$ bins for BDT$_\textrm{ET}$ ({\color{blue}thick} curve) and $\ET^\textrm{prim}$ ({\color{red}thin} curve). Right: Distribution of calibrated $\ET$ resolutions by dividing by $\hat{R}$, the mode of $R$ in each slice; the bin range is also rescaled by $\hat{R}$. After calibration, the rms at unity and the gain with respect to primitive jets are listed.
  \label{fig:jer}
}
\end{figure}

\begin{figure}[tbp!]
\centering\includegraphics[width=0.995\columnwidth]{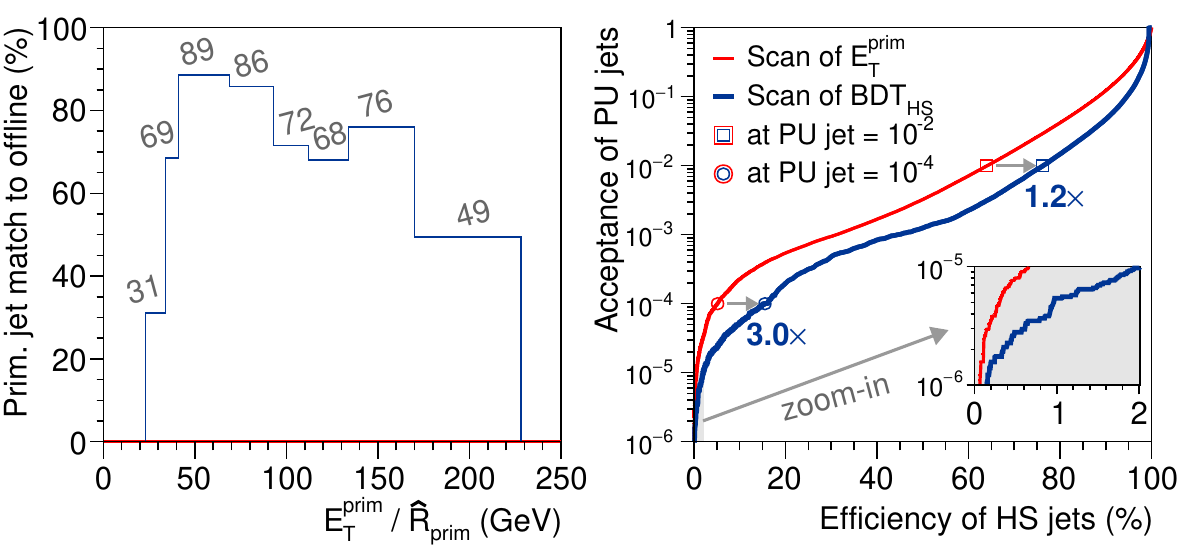}%
\caption{
  Performance of BDT$_\textrm{HS}$. Left: Fraction of primitive jets matched to offline jets in bins of the calibrated $\ET^\textrm{prim}$ (Fig.~\ref{fig:jer}) using the dijet sample. Right: Acceptance of primitive jets identified as PU vs.\ efficiency of primitive jets identified as HS with $\ET^\textrm{offline}{\,>\,}40\,$GeV obtained by scans of BDT$_\textrm{HS}$ ({\color{blue}thick} curve) and $\ET^\textrm{prim}$ ({\color{red}thin} curve) using the $H\!H_{4b}$ sample.
  \label{fig:roc}
}
\end{figure}

We evaluate BDT$_\textrm{ET}$ by considering the ratio $R$ of the regressed $\ET$ and the $\Delta R$-matched reference $\ET^\textrm{offline}$, and is compared to the $R$ for primitive jets. These distributions are shown in bins of $\ET^\textrm{prim}$ in the left plot of Fig.~\ref{fig:jer}. The $R_\textrm{prim}$ distributions are uncalibrated, so they peak at high values of $3.5$ in the lowest $\ET^\textrm{prim}$ bin, as pileup contribution dominate, while it moves towards unity for higher $\ET^\textrm{prim}$ bins as pileup contributions dominate less. In contrast, $R_\textrm{BDT}$ ratios hover around unity for all bins showing the effectiveness of \textit{in situ} pileup correction. The modes of the $R$ distribution ($\hat{R}$) are listed in the left plot and quantifies the amount of pileup in each bin.

The comparison of uncalibrated $\ET$ is made to show the effect of pileup as it does not reflect the performance. To compare the performance, we calibrate each incoming $\ET$ by the $\hat{R}$ derived above for each bin in $\ET^\textrm{prim}$. The resulting distribution centers around unity as shown on the right plot of Fig.~\ref{fig:jer}. The width of the calibrated distributions are listed on the plot along with the percentage gain, rms$_\textrm{BDT}$/rms$_\textrm{prim}{\,-\,}1$. The $\ET$ ranges of the bins are rescaled by $\hat{R}_\textrm{prim}$ to give a range closer to the corresponding offline values. In the calibrated $\ET$ bin of $69$-$93\,$GeV, which is in our target range for $H\!H_{4b}$, the rms gain is $33\%$. If advancements of trigger systems would allow acceptance of lower $\ET$ jets, we note that in the calibrated $\ET$ bin from around $41$-$69\,$GeV, the rms gain is doubled at $57\%$.

Since BDT$_\textrm{ET}$ is designed estimate the $\ET$ from  primitive jets that originate from the HS process, a correction is made to address the fraction not originating from HS. BDT$_\textrm{HS}$ is a classifier trained to categorize primitive jet into HS and PU by utilizing differences in the $\ET$ distribution in each ring. It is trained on the sample of primitive jets using the same set of input features with matched jet given a score of $1$ and unmatched jet given a score of $0$. The output score of BDT$_\textrm{HS}$ is a continuous value between $0$ and $1$. As expected, the likelihood a a primitive jet originates from HS increases with its calorimeter energy, $\ET^\textrm{prim}$. The left plot in Fig.~\ref{fig:roc} shows that after calibration in the corresponding bin as in the right plot of Fig.~\ref{fig:jer}, primitive jets with calibrated $\ET^\textrm{prim}$ in range of interest from $60$-$100\,$GeV are likely to originate from HS at $70$-$90\%$. The effectiveness of classifying primitive jets as HS using BDT$_\textrm{HS}$ is compared to using $\ET^\textrm{prim}$ alone is shown by comparing their ROC curves in the right plot of Fig.~\ref{fig:roc}. At $1\%$ ($0.01\%$) acceptance of PU jets, the improvement in efficiency in identifying primitive jets as HS using BDT$_\textrm{HS}$ is a factor of $1.2$ ($3.0$) higher than just using primitive jets.

We transform the BDT$_\textrm{HS}$ score to a ``HS probability'' by normalizing to the total number of jets, $p_\textrm{HS}{\,=\,}N_\textrm{HS}/(N_\textrm{HS}{\,+\,}N_\textrm{PU})$, where $N_\textrm{HS}$ represents the number of HS events in the bin containing the score in the distribution and likewise for $N_\textrm{PU}$. This probability is used to combine with the output of BDT$_\textrm{ET}$.

Following conventional jet algorithms that use $\ET$ thresholds to trigger events to keep~\cite{atlas-trigger-menu-2018}, the final output of the ML is the estimated BDT$_\textrm{ET}$ value weighted by the HS probability, $p_\textrm{HS}$. The weighting prioritizes those primitive jets that are likely from HS and to suppress those that are likely from stochastic pileup:
\begin{equation}
  \ET^\textrm{ML} = p_\textrm{HS} \cdot \textrm{BDT}_\textrm{ET}.
\end{equation}
The workflow for the real-time inference is shown on the bottom row of Fig.~\ref{fig:flowchart}. The pink box encloses the ML proposed in this paper.

Lastly, the ML must be able to be fit within the constraints of the LHC experiments. Previous work have demonstrated that models with parameters similar to those in our paper can be implemented using percent-level resources with $\mathcal{O}(10)\,$ns latency and interval between successive inputs on FPGA \cite{Hong:2021snb,Carlson:2022dgb,Serhiayenka:2024han,Summers:2020xiy}.

\textit{Results.}---The performance is evaluated using $H\!H_{4b}$ as the physics process of interest. The impact of $H\!H_{4b}$ signal acceptance is demonstrated by comparing equal-background-rate efficiency curves constructed using offline jets as reference. The threshold goal for the HL-LHC four-jet trigger at $80\%$ efficiency is set to $90\,$GeV, consistent with ATLAS in Run 3~\cite{ATLAS:2024xna}. This threshold corresponds to a dijet background acceptance of around $1\%$, which is used to derive the equal-rate threshold the $\ET^\textrm{prim}$-based trigger. The left plot of Fig.~\ref{fig:turnon} shows the distribution of the $\ET^\textrm{offline}$ of the fourth leading jet for $H\!H_{4b}$ events. The distribution peaks around $50\,$GeV and rapidly falls at higher values. In particular, the inset shows the region of interest from $80$-$120\,$GeV along with the cumulative signal fraction, where we see increases by a factor of $1.5$ every $10\,$GeV.

The right plot of Fig.~\ref{fig:turnon} shows the turn-on curve showing $H\!H_{4b}$ signal efficiency as a function of $\ET^\textrm{offline}$ for the fourth leading jet. The plot shows four curves: two sets, for $5\%$ and $1\%$ dijet acceptance, of a pair of curves for primitive jets and ML jets. We see that the two ML curves becomes fully efficient much quicker than their primitive counterparts. A horizontal line at $80\%$ shows $13.5\,$GeV shift from primitive jets to ML jets for $5\%$ dijet acceptance, and $14.5\,$GeV shift for $1\%$ dijet acceptance. To compare with existing values, we consider improvement at $80\%$ efficiency. The improvement using ML reduces the threshold from around $90$ to $75\,$GeV there, increasing the signal acceptance by a factor of $2.1$. In order to achieve a similar gain without the ML, the background acceptance would need to be increased by a factor of $5$. This is demonstrated by the overlap of the ``$5\%$, Prim.\ jets'' curve and ``$1\%$, ML jets'' curve at the horizontal $80\%$ efficiency line. ATLAS projections show that the upper bound on the Higgs self-coupling by lowering the jet threshold improves by a relative $20\%$~\cite{ATLAS:2018rvj}, although this is without systematic uncertainties so the actual improvement is likely to be lower.

\textit{Conclusions.}---As limitations on the minimum $\ET$ threshold primarily occur in the L1 trigger, the jet $\ET$ calibration using ML proposed in this paper for the first-level trigger system can lower the fourth leading jet $\ET$ by $10$-$15\,$GeV, depending on the starting trigger threshold, for $H\!H{\,\rightarrow\,}b\bar{b}b\bar{b}$. This, in turn, may help improve the sensitivity of Higgs self-coupling by up to $20\%$.

\section*{Acknowledgments}
We thank Ariel Schwartzman for suggesting ring$_4$ outside the primitive jet. We thank Steven Schramm pointing out the interplay between classification and regression. We thank Santiago Can\'{e} for FPGA estimations for our models. TMH is supported by the US Department of Energy [award no.\ DE-SC0007914]. BTC is supported by the National Science Foundation [award no.\ 2209370]. STR was supported by the Emil Sanielevici Scholarship.

\FloatBarrier


\begin{thebibliography}{99}
\bibitem{sliding_window}
Algorithm in four steps: generate seeds, requre local maxima, apply threshold, and find jet centers. (1) Each $0.1{\times}0.1$ tower generates a $0.3{\times}0.3$ square ``seed'' from the neighboring towers in each direction, whose sum $\ET$ is the $\ET^\textrm{seed}$. (2) Each seed is compared to the seed $\ET$ of the other 24 seeds centered within a $0.5{\times}0.5$ square around it. Only keep seeds that has the greatest $\ET^\textrm{seed}$ among its neighbors. (3) Discard seeds with $\ET^\textrm{seed}{\,<\,}20\,$GeV. (4) The remaining seeds are the centers of the primitive jets.
\end{thebibliography}
\end{document}